\documentclass[pra,aps,12pt]{revtex4}
\usepackage{amsmath,amsfonts, epsfig, oldgerm,eufrak, graphics}

\newcommand{\ket}[1]{| #1 \rangle}

\newcommand\beq{\begin{equation}}
\newcommand\eeq{\end{equation}}
\newcommand\bea{\begin{eqnarray}}
\newcommand\eea{\end{eqnarray}}
\newtheorem{definition}{Definition}[section]

\newtheorem{theorem}[definition]{Theorem}

\newtheorem{problem}[definition]{Problem}

\begin{document}

\title{Is Entanglement Monogamous?}
\author{Barbara M. Terhal}
%\affiliation{\vspace*{1.2ex}
%            \hspace*{0.5ex}{IBM Watson Research Center,
%P.O. Box 218, Yorktown Heights, NY 10598, USA}}

\address{\vspace*{1.2ex}
            \hspace*{0.5ex}{IBM Watson Research Center,
P.O. Box 218, Yorktown Heights, NY 10598, USA}}

\date{\today}
\begin{abstract}
In this article I discuss Charlie's Bennett's influence in quantum information theory and our answers to the question whether quantum entanglement is
`monogamous'.
\end{abstract}

\maketitle

% close bracket for pretty quant-ph mode
% ]

\section{}
\section*{Charlie's Angels}

It is a pleasure to contribute to the scientific festivities
around Charlie Bennett's 60th birthday. I believe that I am here representing the younger generation of quantum information theorists.
Like John Smolin and Ashish Thapliyal who did also their PhD work at IBM, I can say that Charlie Bennett is my `scientific father'.
Not being at a university, Charlie has never assembled a large group of PhD students. Nonetheless, I think that
Charlie has inspired many of us, students, postdocs and also senior researchers, some of whom are
here at this symposium today, in quantum information or just plainly for the scientific enterprise
itself. In some sense I would say we are all students of Charlie.

\begin{figure*}
\begin{center}
\caption{Charlie's Angels. Participants at the QIP conferences in (1) {\AA}rhus in 1998, (2) Montr\'{e}al in 2000, (3) Amsterdam in 2001, and (4) at IBM Watson in 2002.}
\end{center}
\label{chbangels}
\end{figure*}

Charlie's influence in quantum information theory does not limit
itself to the direct effects of his scientific work, quantum
teleportation, quantum key distribution, the framework of
entanglement manipulations to name a few. There seems to be a deeper
thread running through quantum information theory to which Charlie
has contributed considerably.

The fact is that quantum information is not a subject
that came about yesterday, or after 1994 when Shor discovered
that you could factor large numbers efficiently on a quantum
computer. It is quite a bit older than that. For example, a few years ago I
found a paper on the subject written in 1976 by the Polish
mathematical physicist Roman Ingarden \cite{ingarden:qit}. The title of the paper is "Quantum Information Theory"
and the first few sentences of the abstract go like this:

\begin{quote}
``A conceptual analysis of the classical information theory of Shannon (1948) shows that this theory cannot be
directly generalized to the usual quantum case. The reason is that in the usual quantum mechanics of closed
systems there is no general concept of joint and conditional probability. Using, however, ..., it is possible to construct
a quantum information theory being then a straightforward generalization of Shannon's theory. "
\end{quote}
 This paper was
published in the Polish journal Reports on Mathematical Physics.
I found this paper in the library here at the Watson
Research Center a few years ago. (Actually when I wanted to have
a look at this paper again before this talk, it turned out that
the library did not have the journal it anymore; it has been
unfortunately thrown out after some recent clean-up!)

Ingarden was one of a group
of physicists, many of them in Eastern Europe and the former Soviet Union, who
were thinking about the intersection of information theory,
probability and quantum physics in the 60s and 70s. The most notable in this group
may be Alexander Holevo from the Steklov Mathematical Institute in Russia who is
still very active in modern quantum information theory. His 1973 result that $n$ quantum bits,
or qubits for short, cannot carry more than $n$ bits of
information is frequently invoked in the modern quantum
information theory.

When I saw Ingarden's paper in the journal, I was curious to learn
what it was about. His paper is one of the first attempts at trying to
build a theory of quantum information analogous to Shannon's classical theory of information.
He considers the problem of transmitting classical information through a quantum channel where the output of the
quantum channel is measured by {\em fixed} single-letter projective measurements. Since input and output are then classical,
Shannon's expression of the classical capacity can be directly applied. \\

So now it is an interesting question to consider in what way modern quantum information
theory is different from the earlier work. And here is, I believe, where Charlie comes in.

First there is of course the difference in volume. A search at the quantum physics archive at
${\tt ArXiv.org}$ reveals that over the last years there have appeared about
883 papers with `information' in the Abstract and about 597 papers with 'entanglement'
in the Abstract and this probably understates the current output.

The second difference however is more fundamental. In the modern theory of quantum information we speak
 of quantum systems in the hands of Alice and Bob and Eve --these names first made their appearance in
 cryptography in the seventies--. We say things such as "Alice sends Bob a qubit and forgets what she did",
 "Bob does a measurement and tells Alice", "Eve does a random unitary transformation on her half of the EPR pair", etc.
Even though this seems very casual, it is, at least for the reader introduced to the quantum information
jargon, crystal-clear what is being meant in these situations.

In the earlier work we are more likely to see descriptions such this one from a paper by G. Lindblad in 1973 \cite{lindblad:meas}

\begin{quote}
The map $T_A\colon W \rightarrow W'$  extended to $B(H)$ is a special case of an
expectation in the operator algebra $B(H)$ i.e. a linear map from $B(H)$ into a von
Neumann subalgebra satisfying $T \circ I=I$ where $I$ is the identity operator.
\end{quote}

This example illustrates the fact that the earlier work
on the subject was first of all grounded in mathematical physics and less well connected to a world in which we
may actually send quantum systems around, do measurements and transform states. What has changed since
the sixties and seventies is not that we have actually seen Eve do a random unitary transformation on her half of an EPR
pair. But somehow we have started to dream and imagine that it would happen some day.
Somehow, people at a practical place like IBM started to think about irreversibility, physics
and information. And those people, most in particular Charles Bennett, preferred to think about quantum
information in a more conceptual intuitive way. Now the goal was to ask simple questions which
could have profound nonintuitive answers --think about teleportation--, and in that manner we started to
understand in what way quantum information is different from classical information.

The new language in which the questions were posed was one of action and operation. A prime example is
Charlie's question "What happens when we throw in an EPR pair?" in the discussion that led to the discovery of quantum teleportation.
This operational point of view, i.e. asking how quantum information and entanglement can be used and manipulated for quantum information
processing, has been extremely fruitful and I believe that Charlie Bennett has played a key role in its success.

\section{}
\section*{Is Entanglement Monogamous?}

Let us now turn to the title of this talk and consult our modern-day oracle, the search engine Google.
When I googled 'entanglement monogamous' I got 215 hits in 0.02 seconds and, as you may expect, a fair number of them were
unrelated to science. Here are the first three:
\begin{itemize}
\item "... You can't become entangled simply by talking on the telephone. Entanglement is monogamous
- the more entangled Bob is with Alice, the less entangled he can be ..." at
{\tt qpip-server.tcs.tifr.res.in/\~\,qpip/HTML/Courses/Bennett/TIFR2.pdf}
\item "... this idea doesn't fit with the traditional view of monogamous societies, Siva ... The
technique has the added bonus of improving the entanglement of pairs that pass ..." at
{\tt www.dhushara.com/book/upd3/2002a/28apr01/nsapr.htm}
\item "... My monogamous wonderful forever relationship has fallen apart ... are trying a polyamorous
relationship but it will be long distance and with minimal entanglement...." at
{\tt www.cavegirl.org/polyhell.html}
\end{itemize}

The first quote is taken from a set of lectures that Bennett gave at the Tata research institute in India.
These few phrases suggest exactly what I mean with the question `Is entanglement monogamous?'. Is entanglement
indeed a property that one can only have with one person or quantum system, or can one share it with many?
I would like to consider an example taken from political life. There are two opinions that were considered relevant a few
months ago. Either a person was pro 'U.S. invasion in Iraq' or he/she was against. It is clear that
it did not take much for the president to convince most people in the United States to be pro-war. And
if president Bush had been against the war, then quite likely the American people would all be against the war.
In this sense a classical bit of correlation --to be or not be against the war in Iraq-- can {\em at least
in principle} be shared by an unlimited number of people.

But now consider a scenario in which Bush and Rumsfeld start out in an entangled state containing both opinions:
\begin{equation}
\frac{1}{\sqrt{2}}\left(\ket{{\tt PRO}\hspace{0.2cm}{\tt WAR}}_{\tt Bush} \otimes \ket{{\tt PRO}\hspace{0.2cm}{\tt WAR}}_{\tt Rummy}+\ket{{\tt AGAINST}\hspace{0.2cm}{\tt WAR}}_{\tt Bush} \otimes \ket{{\tt AGAINST}\hspace{0.2cm}{\tt WAR}}_{\tt Rummy}\right)
\end{equation}
i.e. a superposition of two states: One is the state in which Bush and Rumsfeld are both pro-war
and one is a state in which both, unlikely though it may seem, are against the war. In this example the amplitudes for
both states are equal, but we may adjust them, say we assign an arbitrary small amplitude of $\sqrt{\epsilon}$
to the no-war state, to closer mimic the real situation. When both opinions are measured, we find that the
guys always agree and the purely classical correlation that is found can again be shared freely.

The Bush-Rumsfeld state is a pure entangled state (I am taking the liberty to follow
current trends of renaming well-established concepts for political reasons and {\em not} call this an EPR pair). This implies that there is no quantum state shared by three
parties, Bush, Rumsfeld and someone else, say, Joe Smith, such that when we remove Joe we get the Bush-Rumsfeld state
{\em and} at the same time, the total three-party state would not change if we interchange Joe and Rummy.
The reason is simple. Let's represent pro-war by a 0 and against the war by a 1.
The first requirement implies that since the Bush-Rumsfeld state is pure, the state for three parties must be
a state for Joe alone in tensor product with the Bush-Rumsfeld state:
\beq
\ket{{\tt Joe's}\,{\tt state}} \otimes \frac{1}{\sqrt{2}}(\ket{0}_{\tt Bush} \otimes \ket{0}_{\tt Rummy}+\ket{1}_{\tt Bush} \otimes \ket{1}_{\tt Rummy})
\eeq
But then it is clear that there is no
symmetry between Joe and Rummy, Joe is unentangled with Bush whereas Rumsfeld is maximally entangled.
This simple observation is the basis for Bennett's phrase "Entanglement is monogamous", unlike partners
or opinions, entanglement cannot be freely shared.

So now here is a question a modern quantum information theorist typically would ask him/herself. First of all, what
is this observation good for, and secondly, is it true for all entangled states?

It turns out that this property of entanglement is essential in
quantum cryptography. As you know, in 1984 Bennett and Brassard proposed a protocol for
quantum key distribution, a way of letting two people, Alice and Bob, share a set of random perfectly
correlated bits about which no else has any or hardly any information. In other words, Alice and Bob
want to establish `a monogamous correlation'.

Now given the arguments above, it is clear that if Alice and
Bob were to share an entangled state, like the Bush-Rumsfeld state, they would be finished. They'd measure and get the same random bit and no one would know what they got.

Following this line of reasoning, Lo and Chau showed in 1999 \cite{LC:qkdsecurity} how Alice and Bob
can proceed in order to obtain (something close to) a set of entangled states and
in this way they proved the security of an entanglement-based quantum key distribution scheme.
The work by Lo and Chau was the stepping stone on which Shor and Preskill \cite{SP:qkd} in 2000 built their security proof of
Bennett $\&$ Brassard's '84 scheme.

The relation between establishment of secrecy or `monogamous' correlation between parties and the transmission of
quantum information or entanglement has been recently investigated by Igor Devetak at IBM and A. Winter in Bristol.
The basic idea is that coherent versions of schemes to establish secret random bits
lead to optimal protocols that achieve the quantum capacity of a quantum channel \cite{devetak:secret}
or lead to optimal 1-way entanglement distillation protocols \cite{DW:secret}.

\section*{Shareability for general states}

Let us now turn to the second question. Are all entangled states monogamous, more precisely, are mixed entangled
states necessarily monogamous? A first example of a mixed entangled state that is is not monogamous but {\em shareable}
--the term is from Ben Schumacher--, can be found in the 1996 paper by Bennett {et al.}
``Mixed State Entanglement and Quantum Error Correction"  \cite{bennett+:bdsw}.  A noisy quantum channel is constructed in the following way:
With probability $1/2$ the input to a qubit channel is transmitted unchanged to Bob. In that case the eavesdropper Eve has a completely random qubit.
And with probability $1/2$ Eve gets the qubit that Alice sends and Bob gets a completely random qubit. If Alice sends half of a maximally
entangled state to Bob, then one can show that the state for Alice and Bob by itself is still entangled. At the
same time the total state that includes Eve's part is always symmetric with respect to Eve and Bob and so Eve is equally
entangled with Alice. This is an example of shareable noisy entanglement. The shareability directly implies that the 1-way distillable entanglement
of the state is zero and similarly the secret key that can be distilled by 1-way communication is zero.

It is not hard to find other examples of such states and in a recent paper \cite{TDS:bell} we have done so. One considers the following
optimization problem.
\begin{problem}
Given $\rho$ on ${\cal H}_A \otimes {\cal H}_{B_1}$. Is there a symmetric extension of $\rho$ to ${\cal H}_A \otimes {\cal H}_{B_1} \otimes {\cal H}_{B_2}$
such that
\beq
{\rm Tr}_{B_2} \rho'=\rho, \;\;{\rm Tr}_{B_1} \rho'=\rho.
\eeq
\end{problem}

Under some reformulation this optimization problem can be written as a semi-definite
program which either returns a symmetric extension $\rho'$ or returns the answer that there are no feasible solutions for the program
which implies that there is no symmetric extension. One can generalize the problem to multiple parties; one
requires that $\rho$ be symmetrically extendible to systems $B_2,\ldots,B_n$ in the sense that the extension
$\rho'$ be invariant under any permutation of the parties $B_1, \ldots, B_n$. A weaker symmetry requirement on $\rho'$ is
one that says that the original density matrix $\rho$ should equal the reduced density matrix $\rho_{AB_i}$ derived from $\rho'$
for all $i$. All features discussed in the next section hold for both notions of symmetric extension.

The last but perhaps most interesting property of shareable mixed entanglement lies its relation to violations of
Bell's inequalities. At least this is how I started to think about this notion.

\section*{Bell inequalities}

In the sixties John Bell formulated an inequality that has to be obeyed by any theory that is classical and local \cite{bell:epr}.
As it turns out, local measurements on entangled quantum states may violate his inequality. In particular
we know that for every pure entangled state there exists a Bell inequality that is violated.

But as with the notion of shareability, one may also ask whether mixed entangled states violate Bell inequalities. This question
has some history, starting with work by Reinhard Werner showing that there exist states, now called Werner states
which are entangled but do not violate Bell inequalities for any number of local measurement settings \cite{werner:lhv}.
But despite the large body of work of Bell inequalities no clear-cut criteria have been developed to decide whether
a state does or does not violate some Bell inequality.

The problem is first that in order to find a violation a possible infinite number of settings and measurement choices
should be considered, which is not feasible. Secondly it is hard computational problem to enumerate all Bell inequalities for
a given setting; it corresponds to enumerating the facets of some highly symmetric polytope.

In the other direction, there have been no constructive methods for formulating local hidden variable
models for states, if they would exist. And such a thing would be quite desirable. As it turns out there is a very nice relation between the shareability
of entanglement and the existence of a local hidden variable model. The correspondence is the following:

\begin{theorem} \cite{TDS:bell}
Let $\rho$ be a density matrix on a Hilbert space ${\cal H}_A \otimes {\cal H}_{B_1}$. If $\rho$ has a symmetric extension $\rho'$ on
${\cal H}_A \otimes {\cal H}_{B_1} \otimes \ldots \otimes {\cal H}_{B_m}$ then there exists a local hidden variable description of $\rho$ when
Alice has an arbitrary number and Bob has m possible measurements.
\end{theorem}

The intuitive picture behind this theorem is simple. The proof of the theorem rests on three observations. If $\rho$ has a symmetric
extension $\rho'$ then Bob may do his measurements ${\cal M}_{B_1}, \ldots, {\cal M}_{B_m}$ on $\rho'$, that is, he does measurement
${\cal M}_{B_i}$ on the space ${\cal H}_{B_i}$. Since $\rho'$ is a symmetric extension of $\rho$, the joint probabilities of outcome
for some ${\cal M}_{A_j}$ and ${\cal M}_{B_i}$ are the same as for $\rho$. Secondly, Bob's $m$ measurements on $\rho'$ can be viewed as one large
measurement ${\cal M}_{B_1} \times {\cal M}_{B_2} \times \ldots \times {\cal M}_{B_m}$. Thirdly, it is known that there always exist
a local hidden variable description of measurements on a quantum state when one of the two parties has only one measurement.
Therefore the measurements on $\rho'$ have a local hidden variable description from which we can deduce the local hidden variable description
of the original measurements on $\rho$.

Before I finish I would like to come back to the question of my talk and say that Charlie was right --as usual--
that all entanglement is monogamous {\em in an asymptotic sense.} An entangled mixed state may have extensions to
some number of parties but are there entangled mixed states that can be extended to a infinite number of parties?
The answer is no. There is a theorem that says that only unentangled states have infinite symmetric extensions:

\begin{theorem}[Fannes-Lewis-Raggio-Schumacher-Verbeure-Werner]\cite{FLV:sym,RW:meanfield}
A quantum state $\rho$ on ${\cal H}_A \otimes {\cal H}_{B_1}$ is unentangled or separable if and only if
$\rho$ has symmetric extensions $\rho'$ on ${\cal H}_A \otimes {\cal H}_{B_1} \otimes \ldots \otimes {\cal H}_{B_n}$
for all $n=2,3,\ldots$.
\end{theorem}

% The proof of the Fannes-Lewis-Schumacher-Verbeure-Werner theorem uses a similar theorem about symmetric states, the so-called
% quantum de Finetti theorem \cite{CFS:finetti}. The de Finetti theorem says that if a sequence of states $\rho_2$ on ${\cal H}^{\otimes 2}$ to
% $\rho_n$ on ${\cal H}^{\otimes n}$ is invariant under all permutations of the tensor factors for all $n$, then
% $\rho_n$ must be a convex mixture of states $\sigma_i^{\otimes n}$.

In this talk I hope to have conveyed some of the flavor of the questions and answers in quantum information theory and Charlie's role
in this exciting area of research. Thank you for your attention.

\bibliographystyle{hunsrt}
\bibliography{refs}

\begin{thebibliography}{10}

\bibitem{ingarden:qit}
R.S. Ingarden.
\newblock Quantum information theory.
\newblock {\em Reports in Mathematical Physics}, 10:43--72, 1976.

\bibitem{lindblad:meas}
G.~Lindblad.
\newblock Quantum entropy, information and quantum measurements.
\newblock {\em Comm. Math. Phys.}, 33:305--322, 1973.

\bibitem{LC:qkdsecurity}
H.-K. Lo and H.~F. Chau.
\newblock Unconditional security of quantum key distribution over arbitrarily
  long distances.
\newblock {\em Science}, 283(5410):2050--2056, 1999.
\newblock \url{http://arxiv.gov/abs/quant-ph/9803006}.

\bibitem{SP:qkd}
P.W. Shor and J.~Preskill.
\newblock Simple proof of security of the {BB84} quantum key distribution
  protocol.
\newblock {\em Phys. Rev. Lett.}, 85:441--444, 2000,
  \url{http://arxiv.gov/abs/quant-ph/0003004}.

\bibitem{devetak:secret}
I.~Devetak.
\newblock The private classical information capacity and quantum information
  capacity of a quantum channel.
\newblock 2003, \url{http://arxiv.gov/abs/quant-ph/0304127}.

\bibitem{DW:secret}
I.~Devetak and A.~Winter.
\newblock Distillation of secret key and entanglement from quantum states.
\newblock 2003, \url{http://arxiv.gov/abs/quant-ph/0306078}.

\bibitem{bennett+:bdsw}
C.H. Bennett, D.P. DiVincenzo, J.A. Smolin, and W.K. Wootters.
\newblock Mixed state entanglement and quantum error correction.
\newblock {\em Phys. Rev. A}, 54:3824--3851, 1996,
  \url{http://arxiv.gov/abs/quant-ph/9604024}.

\bibitem{TDS:bell}
B.M. Terhal, A.C. Doherty, and D.~Schwab.
\newblock Local hidden variable theories for quantum states.
\newblock {\em Phys. Rev. Lett.}, 90:157903, 2003.

\bibitem{bell:epr}
J.S. Bell.
\newblock On the {Einstein-Podolsky-Rosen} paradox.
\newblock {\em Physics}, 1:195--200, 1964.

\bibitem{werner:lhv}
R.F. Werner.
\newblock Quantum states with {Einstein-Podolsky-Rosen} correlations admitting
  a hidden-variable model.
\newblock {\em Phys. Rev. A}, 40:4277--4281, 1989.

\bibitem{FLV:sym}
M.~Fannes, J.T. Lewis, and A.~Verbeure.
\newblock Symmetric states of composite systems.
\newblock {\em Lett. Math. Phys.}, 15:255--260, 1988.

\bibitem{RW:meanfield}
G.A. Raggio and R.F. Werner.
\newblock Quantum statistical mechanics of general mean field systems.
\newblock {\em Helvetica Physics Acts.}, 62:980--1003, 1989.

\end{thebibliography}

\end{document}